\documentclass[conference]{IEEEtran}
\IEEEoverridecommandlockouts
\usepackage{cite}
\usepackage{amsmath,amssymb,amsfonts}
\usepackage{algorithmic}
\usepackage{graphicx}
\usepackage{textcomp}
\usepackage{xcolor}
\usepackage{longtable}
\usepackage{booktabs}
\usepackage{graphicx}
\usepackage{lipsum}
\usepackage{tabularx}
\usepackage{adjustbox}
\usepackage{comment}
\usepackage{algorithm}

\makeatletter

\@ifundefined{showcaptionsetup}{}{
 \PassOptionsToPackage{caption=false}{subfig}}
\usepackage{subfig}
\makeatother

\usepackage{eso-pic}

\def\BibTeX{{\rm B\kern-.05em{\sc i\kern-.025em b}\kern-.08em
    T\kern-.1667em\lower.7ex\hbox{E}\kern-.125emX}}

\begin{document}

\title{DREAMS: Decentralized Resource Allocation and Service Management across the Compute Continuum Using Service Affinity}

\author{
\IEEEauthorblockN{Hai Dinh-Tuan}
\IEEEauthorblockA{Service-Centric Networking \\
\textit{Technische Universit\"at Berlin}\\
Berlin, Germany \\
hai.dinhtuan@tu-berlin.de}
\and
\IEEEauthorblockN{Tien Hung Nguyen}
\IEEEauthorblockA{
\textit{Technische Universit\"at Berlin}\\
Berlin, Germany \\
nguyen.tienhung@outlook.de}
\and
\IEEEauthorblockN{Sanjeet Raj Pandey}
\IEEEauthorblockA{Service-Centric Networking \\
\textit{Technische Universit\"at Berlin}\\
Berlin, Germany \\
s.pandey@tu-berlin.de}
}

\maketitle

\begin{abstract}

Modern manufacturing systems require adaptive computing infrastructures that can respond to highly dynamic workloads and increasingly customized production demands. The compute continuum emerges as a promising solution, enabling flexible deployment of microservices across distributed, heterogeneous domains. However, this paradigm also requires a novel approach to resource allocation and service placement, as traditional centralized solutions struggle to scale effectively, suffer from latency bottlenecks, and introduce single points of failure. In this paper, we present DREAMS, a decentralized framework that optimizes microservice placement decisions collaboratively across different computational domains. At its core, DREAMS introduces agents that operate autonomously within each domain while coordinating globally through a Raft-based consensus algorithm and cost-benefit voting. This decentralized architecture enables responsive, privacy-preserving, and fault-tolerant coordination, making it particularly suitable given the growing prevalence of multi-stakeholder scenarios across the compute continuum. In particular, within modern manufacturing environments, DREAMS achieves globally optimized service placements while maintaining high fault tolerance. Further evaluations demonstrate that key coordination operations, such as Local Domain Manager (LDM) registration and migration voting, scale sub-linearly with the number of domains, confirming the efficiency and scalability of our proposal.
\end{abstract}


\section{Introduction}

The \textit{Compute Continuum (CC)} is a novel paradigm that unifies cloud, edge, and IoT resources, allowing services to operate seamlessly across centralized and decentralized environments \cite{ullah2023orchestration}. It supports diverse functions such as computation, storage, control, and communication at different layers of the infrastructure. Although the CC is being explored in many domains, its strategic importance is particularly evident in the manufacturing sector, as demonstrated by industrial initiatives such as Audi’s Edge Cloud 4 Production \cite{audi2023edge} and Siemens’ Industrial Edge \cite{beitinger_2019}.

This growing interest is driven by the increasing complexity of manufacturing systems, fueled by emerging technologies such as Internet of Things (IoT), artificial intelligence, and advanced robotics. These innovations have introduced new requirements, including mass customization and supply chain visibility, which were absent in traditional production. Combined with recent global supply chain disruptions and geopolitical uncertainties, manufacturers face increasing pressure to build highly adaptable systems. As a result, there is a growing shift toward paradigms like \textit{Reconfigurable Manufacturing Systems}, which require computing infrastructures for dynamic, decentralized coordination across diverse production settings.

In this paper, we focus on one such use case, where compute resources are distributed across edge devices (for real-time control), regional clouds (for data aggregation), and global clouds (for strategic planning). Services such as scheduling, quality control, and energy optimization are dynamically distributed across the continuum and must be continuously relocated during production to minimize latency and meet Quality of Service (QoS) requirements. To address this, we propose a decentralized approach in which each compute domain autonomously evaluates local metrics and collaboratively decides service placement. Although our framework is evaluated in the context of smart manufacturing, its design is broadly applicable to other multi-domain environments. The main contributions of this work are therefore threefold:
\begin{enumerate}
    \item A decentralized decision-making framework for collaborative resource allocation and service management across the compute continuum.
    \item The design and implementation of a reusable \textit{Local Domain Manager} (LDM), capable of autonomous operation and coordination through consensus mechanisms.
    \item An extensive evaluation demonstrating the feasibility and scalability of the proposed architecture, including metrics as the number of domains increases.
\end{enumerate}

The remainder of this paper is structured as follows: Section II reviews related work, emphasizing their key contributions and limitations. Section III presents the design of our LDM, while Section IV details the technical aspects of the underlying decentralized decision-making logic. Section V describes the evaluation setup and discusses the results. Section VI concludes the paper and outlines directions for future research.

\section{Related work}

Optimizing service placement in distributed environments, particularly for cloud-native services across the CC, is critical for minimizing latency, enhancing resource efficiency, and ensuring robust performance in dynamic, real-time scenarios \cite{mota2024optimizing}. Existing approaches broadly fall into centralized and decentralized categories, each with distinct trade-offs in optimality, scalability, and adaptability.

Traditional \textit{exact methods} guarantee optimal solutions but require global knowledge and incur significant communication and computational overhead. For example, Benamer et al.~\cite{benamer2024semi} modeled microservice placement in hybrid Cloud-Edge/Fog environments as a linear programming (LP) problem, aiming to minimize latency and resource costs under QoS constraints. By combining LP with k-means clustering, the approach significantly reduces execution time in dense and geographically distributed environments. 

\textit{Heuristic strategies} offer computationally efficient alternatives to exact methods, sacrificing optimality for scalability. For instance, Soumplis et al.~\cite{soumplis2024performance} address resource allocation with a Mixed Integer Linear Programming (MILP) model and mitigate its complexity using a greedy best-fit heuristic and a multi-agent rollout, achieving near-optimal results (within 4\%) with faster execution. Another heuristic-based method is SAGA \cite{10577743}, which models microservice interactions as a service affinity graph and applies a modified k-way Kernighan-Lin algorithm to partition services, thereby optimizing placement as a minimum-weight k-cut problem to achieve a 23.40\% reduction in mean latency.

Similarly, \textit{metaheuristic} approaches have been explored to address the complexity of service placement. Wei et al.~\cite{wei2023cost} propose CSRPES, a stochastic algorithm that integrates horizontal and vertical resource sharing to maximize revenue under dynamic demand and nonlinear pricing. By combining fast approximation techniques with Particle Swarm Optimization, Simulated Annealing, and Differential Evolution, the algorithm iteratively refines a base solution and achieves over 30\% revenue gains in resource-constrained environments.

\textit{Machine learning-based approaches} have also recently gained traction. For example, Afachao et al.~\cite{afachao2024efficient} propose BAMPP, a Reinforcement Learning (RL) algorithm for dynamic microservice deployment in CC. Built on Advantage Actor-Critic with enhanced stability via state-value loss ratios, BAMPP reduces network usage (up to 8\%), energy consumption (up to 16.5\%), and migration delays. Similarly, Dinh-Tuan et al. \cite{9751522} propose a novel, self-optimizing microservices system that applies grid and random search techniques to automatically tune runtime configuration parameters, achieving a latency performance improvement of up to 10.56\%.

However, the very nature of the CC, as a unified infrastructure spanning from centralized cloud to edge and IoT devices, implies a massive scale and high degree of distribution. When combined with the proliferation of cloud-native services, these centralized approaches face significant limitations. These include single points of failure, inefficiency in handling large data volumes, and privacy concerns \cite{dinh2024optimizing}. As a result, recent research increasingly favors decentralized approaches for their scalability, resilience, and local autonomy.

One example is EPOS Fog, introduced by Nezami et al.~\cite{nezami2021decentralized}, a decentralized multi-agent framework for IoT service placement across the CC. Each node hosts a local agent that autonomously generates placement plans using local information. Plans are exchanged and optimized via a self-organized tree topology, enabling coordinated decentralized decisions. EPOS Fog reduces execution delay by up to 25\% and improves load balancing by 90\% over centralized and heuristic baselines, showing scalability and resilience in dynamic IoT settings. However, its coordination relies on implicit convergence through cooperative reinforcement learning, which lacks transparency and formal guarantees on consensus, limiting suitability for scenarios requiring explainability, determinism, or multi-stakeholder agreement.

In contrast, DREAMS addresses these gaps with a holistic, context-aware decision model based on \textit{Service Affinity} \cite{dinh2024optimizing}, which captures runtime communication, design-time dependencies, operational patterns, and data privacy requirements. It further introduces a transparent, fair coordination mechanism: a two-phase proposal and voting process with tunable thresholds that evaluates service migrations by both local impact and global benefit. This makes DREAMS well-suited for decentralized, multi-stakeholder environments.

\section{Concept and Design}

At the core of our approach is a multi-agent model, where each computational domain is independently managed and represented by an autonomous \textit{Local Domain Manager (LDM)}. These LDMs monitor local metrics (latency, resource availability, performance, and service interactions) and collaborate to make decentralized service placement decisions. This occurs through a two-phase process: LDMs first propose migrations based on local QoS gains and affinity improvements, then participate in a voting phase to assess local impact and reach consensus. Approved proposals are executed in a rollback-able manner to ensure fault tolerance and consistency. Our design follows these five key principles:

\begin{itemize}
    \item \textit{Self-governed Decentralization:} Each LDM operates autonomously within its domain.
    \item \textit{Privacy-Preserving Computation:} Metrics are processed locally to minimize data exposure.
    \item \textit{Collaborative Optimization:} LDMs coordinate for global goals while respecting local autonomy.
    \item \textit{Heuristic-Driven Placement:} Decisions are driven by an extensible heuristic model for performance, efficiency, and scalability.
    \item \textit{Fault Tolerance:} Coordination is resilient to LDM failures and supports rollback for consistency.
\end{itemize}

\begin{figure}[ht]
    \centering
    \includegraphics[width=\linewidth]{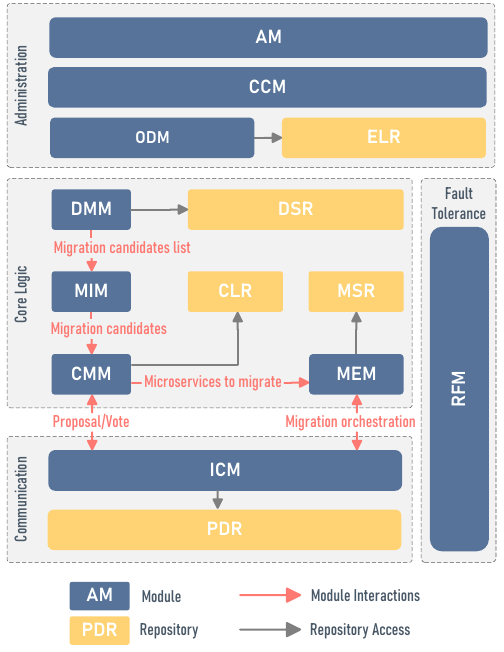}
    \caption{Overview of modules and repositories of the LDM system architecture.}
    \label{fig:ldm-architecture}
\end{figure}

These principles are realized through a \textit{Domain-Driven Design} (DDD) approach that identifies the core domain and decomposes it into bounded contexts, each addressing a specific functional concern. Within these contexts, we define entities, aggregates, and domain services, and use domain events for event-driven communication. Context mapping and integration patterns provide clear boundaries and interaction strategies, supporting modularity and system scalability. The resulting LDM is organized as a composition of modules aligned with these contexts, as illustrated in Figure~\ref{fig:ldm-architecture}.

\noindent\textbf{Administrative Module (AM) - Admin/Policy Context}: Provides tools for administrators to monitor and manage the LDM.
\begin{itemize}
    \item \textit{Administrative Dashboard:} Provides an interface for monitoring local domain's health and migration activity.
    \item \textit{Policy Manager:} Allows administrators to define and update migration policies.
    \item \textit{Visualizer and Report:} Generates visualizations for dependencies, affinities, and performance metrics.
\end{itemize}

\noindent\textbf{Configuration Control Module (CCM) - Configuration Context}: Manages all configuration settings required for the LDM to operate and adapt to changing conditions.
\begin{itemize}
    \item \textit{Configuration Repository Service:} Acts as the central storage and retrieval point for all configuration data.
    \item \textit{Dynamic Configuration Updater:} Manages real-time updates to configuration settings and propagates changes to dependent modules.
    \item \textit{Configuration Validator Service:} Ensures new or updated configurations meet predefined constraints and dependencies.
\end{itemize}

\noindent\textbf{Observability and Diagnostics Module (ODM) - Observability Context}: Monitors the system's performance and provides visibility into operations.
\begin{itemize}
    \item \textit{Metrics Aggregator:} Aggregates metrics for migration success rates and system resource usage.
    \item \textit{Event Logger:} Logs key events for debugging and diagnostics.
\end{itemize}

\noindent\textbf{Domain Monitoring Module (DMM) - Monitoring Context}: Responsible for collecting, normalizing, and maintaining runtime metrics and service interaction data within a local domain. Key domain services are:
    \begin{itemize}
        \item \textit{Service Health Monitor:} Aggregates CPU, memory, and traffic metrics; performs unit conversion, filtering, and outlier removal. Stores data in the Domain State Repository.
        \item \textit{Service Affinity Calculator:} Calculates service affinity scores using communication patterns and resource usage, following the formulas defined in \cite{dinh2024optimizing}. Scores are stored in the Domain State Repository for intra-domain affinities, and in the Peer Domain Repository for inter-domain affinities.
    \end{itemize}

\noindent\textbf{Migration Intelligence Module (MIM) - Optimization/Decision Context}: Manages migration decisions in both leader and follower roles.
    \begin{itemize}
        \item \textit{Migration Eligibility Evaluator:} As leader, it selects migration candidates using service metrics, affinity, latency, and resource data. It uses information from both the Domain State Repository and the Peer Domain Repository to select optimal candidates and forwards them to the Proposal Manager in the Consensus Management Module.
        \item \textit{Cost-Benefit Analyzer:} As follower, it evaluates proposals based on expected QoS, resource, and latency impact to cast a vote.
    \end{itemize} 
    
\noindent\textbf{Consensus Management Module (CMM) - Consensus Context}: Manages decentralized agreement on migration proposals via a Raft-based protocol, ensuring system-wide consistency through quorum and log replication.
\begin{itemize}
        \item \textit{Proposal Manager:} Assembles migration proposals based on MIM inputs.
        \item \textit{Consensus Voting Engine:} Distributes proposals, collects votes, and triggers replication through the Consensus Log Service.
        \item \textit{Leader Coordinator:} Oversees leader election and term management.
        \item \textit{Consensus Fault Recovery:} Handles incomplete consensus and restores state from logs.
    \item \textit{Consensus Log Service:} Stores consensus-related logs in the Consensus Log Repository.
\end{itemize}

\noindent\textbf{Migration Execution Module (MEM) - Execution Context}: Executes and validates service migrations with rollback mechanisms to ensure consistency.
    \begin{itemize}
        \item \textit{Migration Orchestrator:} Executes migrations post-consensus, checking resource availability and coordinating rollback if needed.
        \item \textit{Rollback Manager:} Performs compensating transactions for failed migrations.
        \item \textit{Health Assurance Validator:} Verifies service health before and after migration.
    \end{itemize}

\noindent\textbf{Inter-Domain Communication Module (ICM) - Communication Context}: Enables inter-LDM communication and state awareness.
    \begin{itemize}
        \item \textit{Inter-domain Monitoring and Health Exchange Service (IMHES):} Periodically measures latency and exchanges availability and migration readiness data with peer LDMs. Data is stored in the Peer Domain Repository.
        \item \textit{Inter-domain Migration Coordinator:} Coordinates migration-specific messaging with remote LDMs.
        \item \textit{LDM Discovery Service:} Handles registration and discovery of LDMs, updating their records in the Peer Domain Repository.
    \item \textit{Inter-domain Communication Service:} Provides end-to-end communication channels among LDMs.
\end{itemize}

\noindent\textbf{Recovery and Fault Tolerance Module (RFM) - Recovery Context}: Ensures resilience and quick recovery in case of crashes or failures.
\begin{itemize}
    \item \textit{State Persistence Engine:} Logs migration actions in a fault-tolerant store for traceability and recovery.
    \item \textit{Crash Recovery Manager:} Resumes interrupted processes after crashes.
    \item \textit{Checkpoint Manager:} Saves periodic checkpoints for faster recovery.
\end{itemize}

These modules persist data to dedicated repositories, each of which serves a specific role in maintaining either local or distributed system state. The system includes five such repositories:

\noindent\textbf{Event Log Repository (ELR)} provides observability and diagnostics by storing logs of system health metrics (CPU, memory, and network usage), migration events (successes and failures), error traces, and historical service performance data. It is primarily populated by the ODM's \textit{Event Logger}.

\noindent\textbf{Domain State Repository (DSR)} maintains localized information about the domain. It stores intra-domain service affinity scores, topology, resource availability, and configuration settings. This repository is primarily populated and updated by the services of the DMM.

\noindent\textbf{Consensus Log Repository (CLR)} ensures distributed consistency by storing logs related to the Raft consensus process, including Raft messages, committed migration proposals, leader election history, and replication metadata. Utilized by the CMM, it supports consensus operations and system-wide fault tolerance, ensuring all LDMs maintain a consistent view of agreed actions and enabling recovery from incomplete or failed rounds.

\noindent\textbf{Migration State Repository (MSR)}  tracks the state and progress of service migrations. It stores details about ongoing and completed migrations, including source and target LDMs, progress checkpoints, and the last known execution step. The repository is used and populated by the MEM to ensure that migrations can resume from their last known state in case of failure, prevent conflicting or duplicate migrations, and support rollback operations.

\noindent\textbf{Peer Domain Repository (PDR)} captures information about other LDMs in the system, including their health status, availability, measured inter-domain latency, membership, and registration details. It also stores affinity scores for inter-domain service interactions. Populated by the ICM's IMHES, the PDR is used by the Migration Eligibility Evaluator and the Cost-Benefit Analyzer within the MIM to select suitable migration candidates and evaluate incoming proposals. This supports decentralized coordination and informed voting in service placement.

\section{Implementation}

The previously described architecture was implemented following the principles of \textit{Hexagonal Architecture}, ensuring a clear separation between domain logic, interfaces, and infrastructure to promote modularity, testability, and maintainability. The LDM is implemented in Java using the lightweight, cloud-native \textit{Quarkus} framework with \textit{GraalVM} support for optimized runtime performance. Inter-module and inter-LDM communication is message-driven and asynchronous, built on \textit{Apache Pekko} (forked from Akka) with clustering and service discovery. Messages are serialized using \textit{Protobuf}. Decentralized consensus is achieved via the \textit{Raft} algorithm, implemented using \textit{Apache Ratis}, supporting leader election, log replication, and agreement on migration proposals. Two in-memory caches using \textit{Caffeine} accelerate access to metrics and service states. The system is deployed on \textit{Kubernetes} (GKE), integrating with its API for resource awareness and service migration.

\subsection{Migration candidate selection}

To identify the optimal microservice for migration, each LDM executes a local decision-making procedure based on affinity analysis and latency evaluation. The process is structured as follows:

\begin{enumerate}
    \item \textbf{Filtering Phase:}  
    Eliminate from consideration all microservices that are:
    \begin{itemize}
        \item explicitly marked as non-migratable, or
        \item already located in the cluster with their highest affinity.
    \end{itemize}

    \item \textbf{Cluster Affinity Computation (\( A_c \)):}  
    For each candidate microservice \( m \), and for each cluster \( c \) in the system, compute the \textit{Cluster Affinity Score}. This score, \( A_c(m) \), represents the total affinity between \( m \) and all the microservices residing within that specific cluster \( c \):
    \[
    A_c(m) = \sum_{v \in \mathcal{N}_c(m)} a_{m,v}(\Delta t)
    \]
    where \( \mathcal{N}_c(m) \) is the set of microservices in cluster \( c \) connected to \( m \), and \( a_{m,v}(\Delta t) \) denotes the affinity between \( m \) and \( v \) over the time window \( \Delta t \). 

    \item \textbf{Affinity Scores:}  
    Define the intra-cluster and inter-cluster affinities as:
    \[
    A_{\text{intra}} = A_{c_{\text{current}}}(m), \quad
    A_{\text{inter}} = \max_{c \ne c_{\text{current}}} A_c(m)
    \]
    where \( A_{\text{intra}} \) represents local affinity within the current cluster and \( A_{\text{inter}} \) denotes the strongest affinity to any other cluster.

    \item \textbf{Affinity Gain (\( \Delta A \)):}  
    Compute the potential gain in affinity from migrating the microservice to the most favorable external cluster:
    \[
    \Delta A = A_{\text{inter}} - A_{\text{intra}}
    \]
    A positive \( \Delta A \) indicates that another cluster has stronger overall affinity to the microservice than its current cluster, suggesting potential QoS improvement through migration.

    \item \textbf{Latency Penalty (\( L \)):}  
    Introduce a penalty for migrating to a more distant cluster, scaled by the affinity gain:
    \[
    L = 
\begin{cases}
\dfrac{\ell}{1 + \exp\left(\dfrac{\Delta A}{\gamma_{\text{proposal}}}\right)}, & \text{if } \Delta A > 0 \\
\ell, & \text{otherwise}
\end{cases}
    \]
    where \( \ell \) is the latency to the target cluster and \( \gamma_{\text{proposal}} \) is the \textit{Affinity Gain Sensitivity Coefficient}.

    \item \textbf{QoS Improvement Score (\( Q \)):}  
    Estimate the benefit of migration using:
    \[
    Q = \Delta A - L = A_{\text{inter}} - A_{\text{intra}} - L
    \]

    \item \textbf{Candidate Selection:}  
    Identify the microservice with the highest \( Q \). If \( Q > \theta_{\text{proposal}} \), where \( \theta_{\text{proposal}} \) is the proposal threshold, the migration is proposed.
\end{enumerate}

This workflow enables each LDM to identify migration candidates based on a balance between improved service affinity and increased communication latency (if any). The decision to propose a migration is governed by two tunable parameters that influence the system's responsiveness and migration selectivity. The \( \gamma_{\text{proposal}} \) shapes the steepness of the sigmoid function used in the latency penalty calculation. A smaller \( \gamma_{\text{proposal}} \) results in a sharper transition, allowing only migrations with significant affinity gains to justify the latency overhead, thereby favoring conservative and stable migration behavior. Conversely, a larger value makes the penalty curve flatter, allowing more aggressive migrations even for modest affinity improvements, which can help the system respond quickly to dynamic changes but may increase churn. The second parameter, the proposal threshold \( \theta_{\text{proposal}} \), acts as a decision boundary for the overall QoS improvement score. It ensures that only candidates with a positive net benefit, taking into account both affinity gains and latency costs, are forwarded for system-wide evaluation. Together, these parameters provide fine-grained control over the proposal process, enabling system designers to adjust the aggressiveness of migration decisions according to application-specific QoS requirements and the stability of the environment.

\subsection{Voting Procedure for Migration Approval}

After the leader LDM proposes a migration, participating LDMs independently evaluate the proposal based on local impact and system-wide trade-offs. The decision-making procedure for each voting LDM proceeds as follows:

\begin{enumerate}
    \item \textbf{Local Impact Score (\( I_{\text{local}} \)):}  
    Assess the dependency of local microservices on the migrating microservice \( m \):
    \[
    I_{\text{local}} = \sum_{v \in \mathcal{N}_{\text{local}}(m)} a_{m,v}(\Delta t)
    \]

    \item \textbf{Non-Affected Clusters:}  
    If \( I_{\text{local}} = 0 \), it is assumed that the migration does not affect local performance, and the LDM casts a \textit{positive vote} immediately.

    \item \textbf{Affinity Normalization (\( \tilde{I}_{\text{local}} \)):}  
    Normalize the local impact:
    \[
    \tilde{I}_{\text{local}} = \frac{I_{\text{local}} - I_{\min}}{I_{\max} - I_{\min} + \epsilon}
    \]
    where \( I_{\min} \) and \( I_{\max} \) are historical extremes, and \( \epsilon \) is a small constant to avoid division by zero.

    \item \textbf{Latency Difference (\( \Delta \ell \)):}  
    Calculate the change in communication latency from the voting LDM's perspective:
    \[
    \Delta \ell = \ell(c_{\text{target}}, c_v) - \ell(c_{\text{source}}, c_v)
    \]

    \item \textbf{Affinity Penalty Weight (\( W_{\text{aff}} \)):}  
    Apply a penalty based on the normalized affinity impact:
    \[
   W_{\text{aff}} = \frac{1}{1 + \exp\left(-\dfrac{\tilde{I}_{\text{local}}}{\gamma_{\text{vote}}}\right)}
    \]

    \item \textbf{Scaled Latency Penalty (\( P_{\text{lat}} \)):}  
    Evaluate the weighted cost of increased latency:
    \[
    P_{\text{lat}} = \frac{\Delta \ell \cdot W_{\text{aff}}}{\ell_{\text{max}}}
    \]

    \item \textbf{Voting Decision:}  
    If \( P_{\text{lat}} < \theta_{\text{vote}} \), the LDM casts a \textit{positive vote}; otherwise, it casts a \textit{negative vote}. This tolerance-based mechanism ensures global gains are not blocked by minor local penalties.
\end{enumerate}

The voting procedure was designed to ensure that migration proposals are evaluated in a decentralized yet coordinated fashion, considering the local impact on each LDM. Two tunable parameters govern this evaluation. The \textit{Local Impact Sensitivity Coefficient} \( \gamma_{\text{vote}} \) controls how sharply an LDM reacts to its own dependency on the migrating service. It adjusts the sigmoid function that weighs the local impact when computing the scaled latency penalty. A smaller value makes LDMs highly sensitive to even small dependencies, effectively favoring local stability. In contrast, a larger \( \gamma_{\text{vote}} \) softens this response, encouraging more global cooperation even at some local cost. The voting threshold \( \theta_{\text{vote}} \) determines the upper limit of the acceptable latency penalty. If the scaled penalty exceeds this threshold, the LDM rejects the migration proposal to preserve local performance. These two parameters allow each LDM to balance its autonomy with collective optimization goals, enabling a flexible and robust consensus process in a multi-stakeholder continuum. The overall design ensures that local evaluations contribute meaningfully to globally beneficial decisions, while still preserving local service quality.

\section{Evaluation}

\subsection{Qualitative evaluation}

DREAMS is designed to support decentralized coordination across a globally distributed compute continuum, where nodes operate independently yet must occasionally reach agreement on critical operations like service migration. This section evaluates the qualitative rationale behind key architectural choices in DREAMS, namely consensus, peer discovery, and consistency model, highlighting how each contributes to scalability, responsiveness, and resilience in heterogeneous, latency-sensitive environments.

\paragraph*{Consensus Mechanism}  
We adopt the \textbf{Raft consensus algorithm} to coordinate migration decisions in a fault-tolerant manner. Although Raft is not Byzantine Fault Tolerant (BFT) and assumes crash-only failures, it maintains an effective balance between simplicity, performance, and correctness in settings where nodes are cooperative but may experience crashes or network partitions. Compared to more complex alternatives like BFT-based protocols or Paxos, Raft offers lower latency, better scalability, and easier implementation—making it well-suited for dynamic, resource-constrained domains in the compute continuum~\cite{kondru2024raftoptima}. Raft tolerates up to $\lfloor (N - 1)/2 \rfloor$ node failures and provides log replication and quorum-based agreement without excessive coordination overhead.

\paragraph*{Peer Discovery and Monitoring}  
Effective decentralized coordination also requires each LDM to discover and monitor peers in real time. Various options exist for this task, including centralized registries (e.g., \texttt{etcd}, \texttt{Consul}), static bootstrapping, distributed hash tables (DHTs), and overlay libraries like \texttt{libp2p}. Centralized solutions simplify discovery but violate decentralization principles and introduce single points of failure. Static configurations are brittle in dynamic environments, while DHTs and overlays increase protocol complexity. To avoid these limitations, DREAMS employs a lightweight \textbf{gossip-based protocol} to disseminate peer health and availability information. This approach provides resilience, scalability, and adaptability under network churn or partial failures, aligning with the system's distributed nature.

\paragraph*{Consistency Model}  
Given the wide-area deployment of LDMs and the varied pace at which domains operate, enforcing strong consistency at all times would significantly degrade responsiveness. Therefore, DREAMS adopts a \textbf{hybrid consistency model}. For non-critical operations such as health monitoring and affinity estimation, modules like the Inter-domain Monitoring and Health Exchange Service (IMHES) and the Service Affinity Calculator operate under eventual consistency. This allows decisions to be made using slightly stale data, which is acceptable for trend-based optimization and improves system responsiveness. In contrast, strong consistency is enforced selectively during migration coordination, where the Raft-based consensus guarantees agreement and system-wide correctness. This targeted enforcement of strong consistency ensures correctness without compromising scalability or latency elsewhere.

\begin{figure}[ht]
    \centering
    \includegraphics[width=\linewidth]{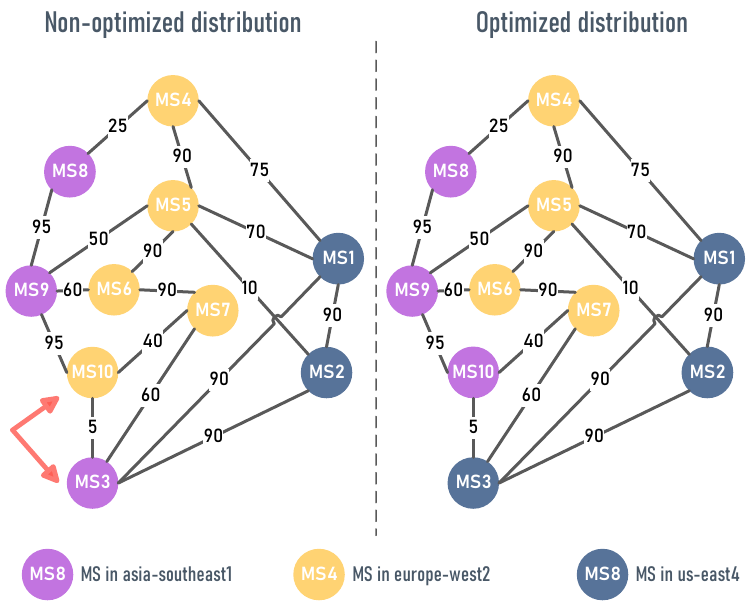}
    \caption{Microservice distribution before (left) and after (right) optimization.}
    \label{fig:before-after}
\end{figure}

To evaluate the system behavior and the global optimization capabilities of the LDMs, we deployed three LDM clusters in geographically distributed Google Cloud regions: \textit{us-east4} (Northern Virginia, USA), \textit{europe-west3} (Frankfurt, Germany), and \textit{asia-southeast1} (Jurong West, Singapore). Each LDM was hosted on an \texttt{e2-standard-4} virtual machine (4 vCPUs, 16\,GB RAM), running Ubuntu 22.04 LTS with identical configurations. Each LDM managed a simulated network of microservices with predefined inter-service affinity values. These affinities were assigned between each pair of services to reflect realistic interaction patterns, as illustrated in Figure~\ref{fig:before-after}. The optimization problem can be formalized as follows: given an undirected graph \( G = (V, E) \) where edge weights \( w(e) \) denote affinity scores, the objective is to minimize the sum of affinities across domain boundaries:

\[
\min_{\substack{u \in V_i,\, v \in V_j \\ i \ne j}} \sum w(u, v)
\]

In the initial (non-optimized) distribution, the total inter-domain affinity value was 630. A globally optimized solution, computed offline, yielded a significantly lower cost of 395. Achieving this optimal configuration involves migrating \texttt{MS3} from \texttt{asia-southeast1} to \texttt{us-east4}, and \texttt{MS10} from \texttt{europe-west3} to \texttt{asia-southeast1}.

During the evaluation, the system demonstrated its capability to converge toward this optimal state. In a normal operational scenario, the LDMs in \texttt{asia-southeast1} and \texttt{europe-west3} successfully identified the correct migration candidates and independently initiated the appropriate proposals. In a fault-tolerance scenario, we simulated the failure of the leader LDM. As expected in a Raft-based system, the cluster correctly entered a brief leader election phase, during which optimization was temporarily suspended as no new proposals could be committed. Once the remaining nodes elected a new leader, the system seamlessly resumed its optimization process. The failed LDM was later able to recover and rejoin the cluster without manual intervention, validating the framework's resilience and self-healing capabilities.

\subsection{Quantitative evaluation}

To complement the qualitative discussion above, we continue to quantitatively evaluate two core procedures within the system. The first is LDM registration, which is handled via gossip-based cluster membership and service discovery mechanisms. The second is the migration voting process, which combines asynchronous message exchange for proposal evaluation with consensus and decision logging through the Raft protocol.

\subsubsection{LDM registration}

To measure LDM registration time, we synchronized clocks across nodes using \texttt{chrony}. Each LDM tracked cluster membership via Pekko’s \texttt{ClusterEvent.MemberUp}. Registration time was defined as the interval between an LDM’s startup and its recognition by all peers. Experiments used cluster sizes of 3–20 LDMs with 2 seed nodes, each repeated three times to obtain average values.

\begin{figure}[ht]
    \centering
    \includegraphics[width=\linewidth]{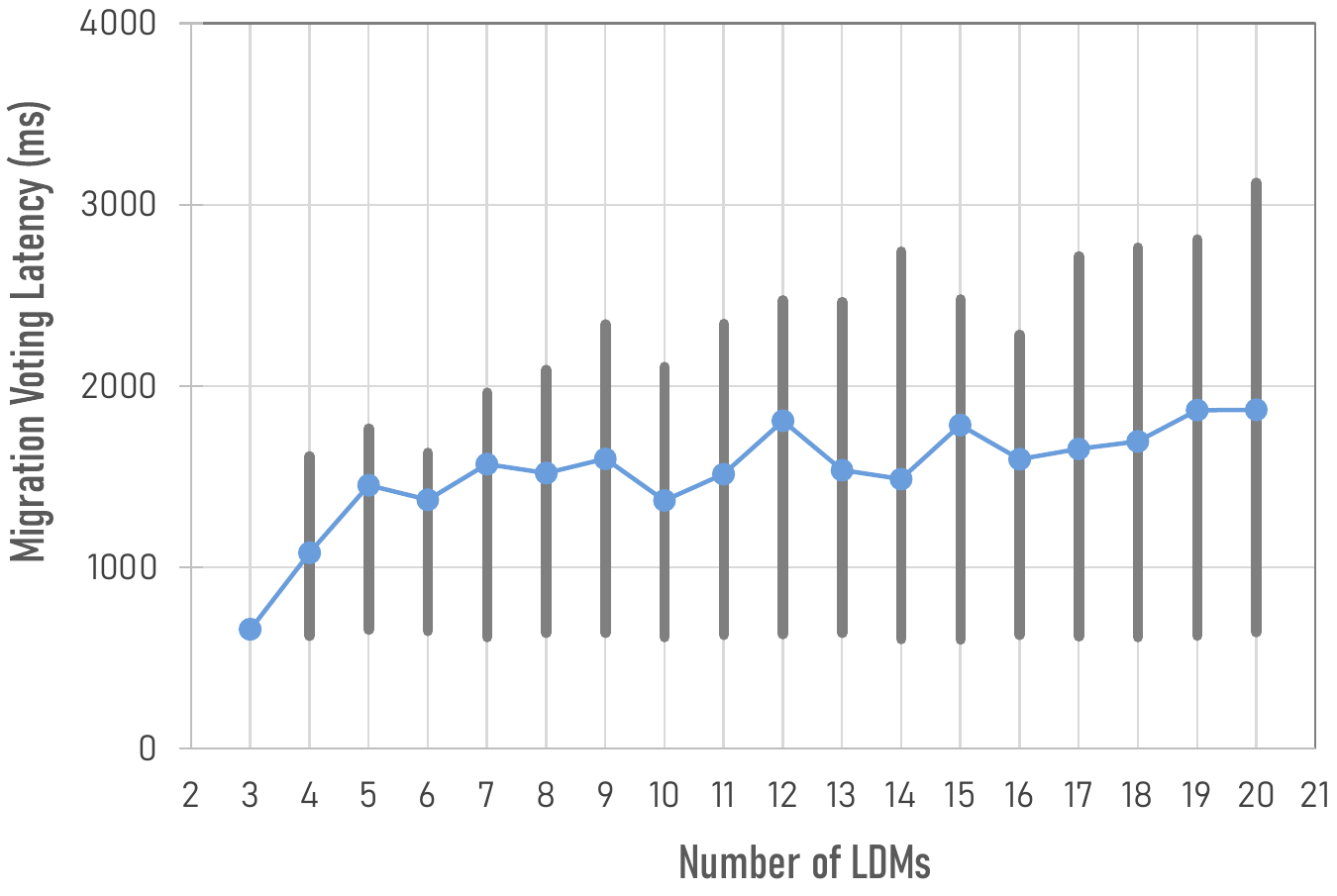}
    \caption{LDM registration latency across cluster sizes.}
    \label{fig:ldm-registration}
\end{figure}

 The results are illustrated in Figure~\ref{fig:ldm-registration}. The vertical bars represent the range between the minimum and maximum registration times observed across all existing LDMs, while the blue line indicates the mean registration time. The minimum registration time (\textit{best-case scenario}) remains relatively stable across all configurations (average 623.9ms), which is expected since the seed nodes, typically the first contacted, acknowledge the new member almost immediately. Indeed, these values are recorded on either one of the two seed nodes. In contrast, the maximum registration time reflects the convergence latency, i.e., the time required for all nodes to recognize the new member. This value increases with the number of LDMs, from 556ms (3 nodes) to 3,121ms (20 nodes). This value represents the \textit{convergence time}, the time it takes for the last or most distant node in the cluster to learn about the new member through the probabilistic gossip protocol. As the cluster grows, the number of random hops the message must take to reach every peer can increase, leading to this higher worst-case latency. Crucially, the mean registration time grows sub-linearly with the number of LDMs, indicating that new members can be integrated efficiently without proportional overhead and that the registration process will not become a bottleneck even in much larger networks.

\subsubsection{Migration voting}

To evaluate the migration voting procedure, we measured the time from proposal initiation by the leader LDM to its commitment in the replicated log. This duration captures the lifecycle of the decentralized coordination process, including (1) the propagation of the proposal to all participating LDMs via Apache Pekko’s asynchronous messaging, (2) the local evaluation and submission of votes based on affinity and latency metrics, and (3) the aggregation of votes by the leader and the subsequent consensus and replication using Apache Ratis. For this evaluation, we omit the actual execution of the migration action, as it is irrelevant to the scalability performance of the decentralized decision-making mechanism. Instead, upon receiving a majority of positive votes, the leader immediately appends the migration decision to the Raft-based replicated log. This approach allows us to isolate and analyze the responsiveness and scalability of the coordination workflow itself, independent of application-specific service migration overhead.

\begin{figure}[ht]
    \centering
    \includegraphics[width=0.93\linewidth]{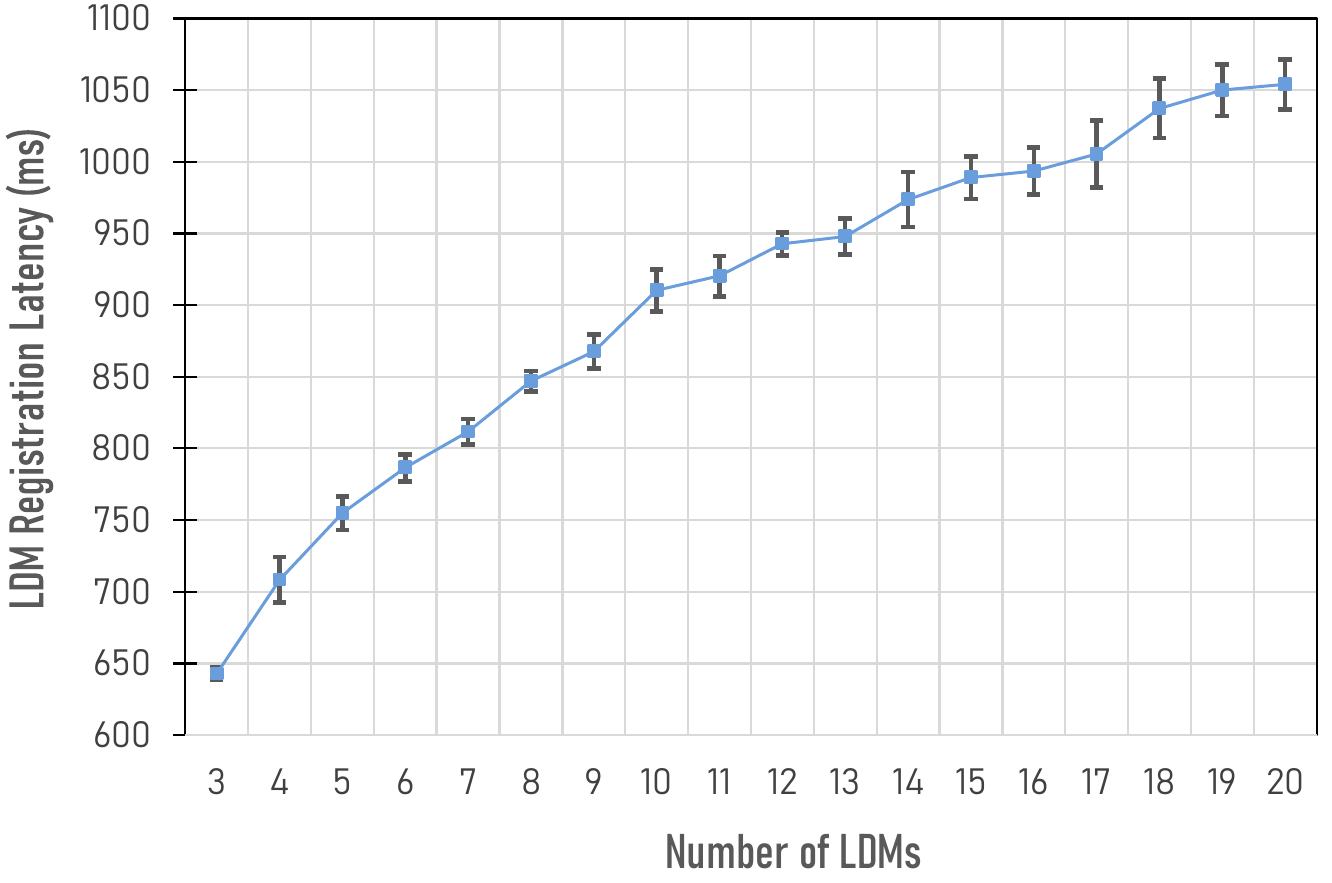}
    \caption{Migration voting latencies across cluster sizes.}
    \label{fig:migration}
\end{figure}

Figure~\ref{fig:migration} illustrates the average time required to complete the migration voting process, measured from the initiation of a proposal by the leader LDM to the point at which the decision is committed through Raft-based log replication. Each data point represents the mean of five independent measurements, and vertical error bars denote the standard deviation.

The results reveal two key findings. First, the low standard deviations across all cluster sizes indicate that the protocol exhibits highly predictable and stable performance. Second, and more importantly, the system demonstrates a clear sub-linear growth trend in latency. The average decision time increases from 642.80 ms with 3 nodes to 1,054.10 ms with 20 nodes, with the curve slightly flattening as the cluster grows. To understand this sub-linear performance, we can analytically decompose the process:

\begin{enumerate}
    \item \textit{Asynchronous Proposal Propagation:}  
    The proposal is broadcast to all $N$ nodes concurrently. Latency is $\mathcal{O}(1)$, bounded by network conditions rather than $N$.

    \item \textit{Parallelized Local Evaluation:}  
    All $N$ LDMs evaluate independently in parallel. The phase completes with the slowest node, giving $\mathcal{O}(1)$ complexity.

    \item \textit{Quorum-Based Vote Aggregation:}  
    The leader stops after a quorum ($N/2+1$) replies. Latency depends on the median voter, yielding $\mathcal{O}(1)$ complexity.

    \item \textit{Constant-Time Raft Commit:}  
    Log entries are sent in parallel and committed once a majority acknowledges, keeping latency independent of $N$ and thus $\mathcal{O}(1)$.
\end{enumerate}

Therefore, the migration voting latency is theoretically $\mathcal{O}(1)$. While this idealized model predicts constant-time performance, our empirical results show a slight upward curve. This is not a contradiction, but an expected reflection of the difference between a theoretical complexity model and a real-world system. The observed sub-linear growth is caused by second-order effects not captured by Big-$\mathcal{O}$ notation, such as a minor increase in workload on the leader node as $N$ grows, and the statistical effect that the time to hear back from the median node of a larger group is slightly longer.

\section{Conclusion \& Future Work}

This paper presents DREAMS, a framework for resource allocation and service management across the CC. 
Building upon prior work introducing the concept of service affinity and a centralized approach (SAGA), DREAMS generalizes this idea into a decentralized and scalable framework that addresses the challenges of dynamic, heterogeneous environments. 
The framework is specifically designed to support coordination across independently managed edge, fog, and cloud domains, often involving multiple stakeholders with diverse operational constraints.

A key contribution of this work is a modular, reusable framework for collaborative service optimization in the CC. Combining affinity-based heuristics with quorum-driven voting and Raft-based consensus, DREAMS enables hybrid consistency and fault-tolerant coordination while preserving data privacy through local processing in autonomous domains. A prototype implementation validates the feasibility of the approach. Experimental results demonstrate that LDMs can collaboratively achieve globally optimized service placements. Quantitative evaluation shows sub-linear scalability in both LDM registration and migration coordination, confirming the architecture's practical efficiency and robustness not only in Industry~4.0 use cases but also in broader settings.

Nonetheless, several limitations remain, offering opportunities for future work. First, the current system assumes cooperative domains and uses a crash-tolerant consensus model. Incorporating Byzantine fault tolerance could enhance robustness in adversarial settings, though at the cost of increased complexity, communication overhead, and governance requirements. Second, the leader election currently uses a simple round-robin mechanism, fair but suboptimal in dynamic environments, where a more adaptive, capability-aware policy could improve responsiveness. Third, the system currently allows only one active migration proposal per leader term; future work could explore batching, prioritization, or parallel proposal handling to better cope with bursts in reconfiguration demands. Finally, parameter tuning (e.g., \( \gamma_{\text{proposal}} \), \( \gamma_{\text{vote}} \), \( \theta_{\text{vote}} \)) could be automated using machine learning to enhance adaptability.

In summary, DREAMS demonstrates that decentralized service management based on service affinity and quorum-based coordination is both feasible and effective for large-scale, multi-domain computing environments. This approach lays a foundation for scalable and resilient industrial CC systems, with potential applications extending to sectors such as healthcare, smart mobility, energy, and beyond.

\bibliographystyle{IEEEtran}
\bibliography{references}

\end{document}